\documentclass[a4paper,fleqn]{cas-dc}

\usepackage[authoryear]{natbib}

\usepackage{amsmath, amsfonts}
\usepackage{graphicx}
\usepackage{booktabs}
\usepackage{multirow}
\usepackage{float}
\usepackage{adjustbox}

\usepackage{algorithm}
\usepackage{algpseudocode}

\usepackage{tikz}
\usepackage{pgfplots}
\usepackage{xcolor}

\usepackage{tcolorbox}
\tcbuselibrary{breakable}

\pgfplotsset{compat=1.18}
\definecolor{gcncolor}{RGB}{211,47,47}
\definecolor{n2cattnTcolor}{RGB}{255,167,38}
\definecolor{n2cattnLcolor}{RGB}{187,156,229}
\definecolor{clusterattncolor}{RGB}{77,182,172}
\definecolor{nodeattncolor}{RGB}{25,79,126}

\definecolor{promptbg}{RGB}{230,230,230}  
\definecolor{prompttitle}{RGB}{64,64,64}   
\definecolor{promptblue}{RGB}{0,100,200}
\definecolor{promptborder}{RGB}{0,0,0}     

\newtcolorbox{codeprompt}[1]{
    colback=promptbg,              
    colframe=promptborder,         
    fonttitle=\bfseries\color{white}\small,
    colbacktitle=prompttitle,      
    title=#1,
    arc=4mm,                       
    outer arc=4mm,                 
    boxrule=1.5pt,                 
    toprule=1.5pt,                 
    bottomrule=1.5pt,              
    leftrule=1.5pt,                
    rightrule=1.5pt,               
    toptitle=1mm,
    bottomtitle=1mm,
    left=5mm,
    right=5mm,
    top=5mm,
    bottom=5mm,
    breakable,
    sharp corners=downhill,        
    rounded corners=all            
}

\begin{document}
\let\WriteBookmarks\relax
\def\floatpagepagefraction{1}
\def\textpagefraction{.001}

\shorttitle{ }

\shortauthors{Zhihong Liang et~al.}

\title [mode = title]{Breaking Obfuscation: Cluster-Aware Graph with LLM-Aided Recovery for Malicious JavaScript Detection}                      
\tnotemark[1]

\tnotetext[1]{This document is the result of a research project supported by the Guangdong Provincial Key Laboratory of Power System Network Security.}

\author[1,2]{Zhihong Liang}
\author[3]{Xin Wang}
\author[3]{Zhenhuang Hu}
\author[3]{Liangliang Song}[%
    orcid=0009-0004-4555-3773]
\author[1,2]{Lin Chen}
\author[3]{Jingjing Guo}
\author[3]{Yanbin Wang}[%
    orcid=0000-0003-1682-5712]
\cormark[1]
\author[3]{Ye Tian}[%
    orcid=0000-0003-0608-8544]
\cormark[1]

\affiliation[1]{organization={Electric Power Research Institute, CSG},
    city={Guangzhou},
    state={Guangdong},
    country={China}}

\affiliation[2]{organization={Guangdong Provincial Key Laboratory of Power System Network Security},
    city={Guangzhou},
    state={Guangdong},
    country={China}}

\affiliation[3]{organization={Hangzhou Institute of Technology, Xidian University},
    city={Hangzhou},
    country={China}}

\cortext[1]{Corresponding author: Yanbin Wang (wangyanbin15@mails.ucas.ac.cn), Ye Tian (tianye@xidian.edu.cn)}


\begin{abstract}
With the rapid expansion of web-based applications and cloud services, malicious JavaScript code continues to pose significant threats to user privacy, system integrity, and enterprise security. However, detecting such threats remains challenging due to sophisticated code obfuscation techniques and JavaScript's inherent language characteristics, particularly its nested closure structures and syntactic flexibility. In this work, we propose DeCoda, a hybrid defense framework that combines large language model (LLM)-based deobfuscation with code graph learning: (1) We first construct a sophisticated prompt-learning pipeline with multi-stage refinement, where the LLM progressively reconstructs the original code structure from obfuscated inputs and then generates normalized Abstract Syntax Tree (AST) representations; (2) In JavaScript ASTs, dynamic typing scatters semantically similar nodes while deeply nested functions fracture scope capturing, introducing structural noise and semantic ambiguity. To address these challenges, we then propose to learn hierarchical code graph representations via a Cluster-wise Graph that synergistically integrates graph transformer network, node clustering, and node-to-cluster attention to simultaneously capture both local node-level semantics and global cluster-induced structural relationships from AST graph. Experimental results demonstrate that our method achieves F1-scores of 94.64\% and 97.71\% on two benchmark datasets, demonstrating absolute improvements of 10.74\% and 13.85\% over state-of-the-art baselines. In false-positive control evaluation at fixed FPR levels (0.0001, 0.001, 0.01), our approach delivers 4.82×, 5.91×, and 2.53× higher TPR respectively compared to the best-performing baseline. These results highlight the effectiveness of LLM-based deobfuscation and underscore the importance of modeling cluster-level relationships in detecting malicious code. Our code is available at the following link: \url{https://github.com/zer0p0intvvv/DeCoda}.
\end{abstract}



\begin{keywords}
Malicious Code Detection \sep JavaScript \sep Graph Neural Networks \sep Deobfuscation \sep Large Language Models
\end{keywords}

\maketitle

\section{Introduction}





JavaScript \cite{10.1145/3386327} is a foundational technology for modern web development, powering dynamic and interactive web applications. However, its widespread adoption and dynamic nature also make it a frequent target for malicious exploitation. Attackers inject obfuscated scripts into vulnerable websites and web services\cite{tian2025webguard++,li2016relationship,malik2019nl2type,lee2023adcpg,liu2025pmanet,liu2024transurl,wang2023large,song2025obfuscated,wang2022snwf,liu2025pmanet}, aiming to steal sensitive data, hijack user sessions, or deploy further payloads. These threats extend beyond traditional web environments, impacting emerging domains \cite{zhang2024hybrid} such as IoT devices, cloud platforms, and even autonomous systems where JavaScript-based interfaces are increasingly utilized. 

To evade detection, these scripts are often obfuscated using techniques \cite{10.1145/2886012,behera2015different} such as variable renaming, control flow distortion, string encoding, and dynamic function calls, which obscure both their syntactic form and semantic intent \cite{skolka2019anything,wei2025detection}. While traditional detection approaches such as signature-based scanning and sequence-based machine learning models (e.g., BERT \cite{koroteev2021bert}, LSTM \cite{greff2016lstm}, and CNN \cite{chua2002cnn}) \cite{sun2021anomaly,zhang2024hybrid} have proven effective against simpler threats, their reliance on linear or localized code representations fundamentally limits their ability to model the hierarchical and relational dependencies essential for analyzing obfuscated code. 

In contrast, Graph Neural Networks (GNNs) \cite{corso2024graph,xiao2025graphedge} provide a distinct paradigm by representing source code as structured graphs, such as abstract syntax trees (ASTs) \cite{yamaguchi2014modeling}. These representations inherently capture syntactic and structural relationships, enabling models to encode dependencies effectively. However, existing methods fail to account for JavaScript-specific language characteristics, such as syntactic flexibility and deep closure nesting structures.

Overall, malicious JavaScript detection faces two critical challenges: (1) The language's flexibility facilitates diverse obfuscation techniques to evade pattern-matching detectors, while current deobfuscation approaches relying on static rule-based transformations cannot handle JavaScript's polymorphic code variations. (2) Prior GNN implementations overlook fundamental JavaScript attributes: (a) they ignore AST node clustering properties - where JavaScript's syntactic flexibility causes semantically equivalent nodes to scatter in feature space, inducing semantic interpretation errors, and (b) their message passing mechanisms disrupt critical scope-chain dependencies, while deeper GNN architectures exacerbate the oversmoothing of closure hierarchies, progressively losing granular scope information across network layers.

To address these challenges, we propose a novel hybrid framework for malicious JavaScript detection that synergizes LLM-guided deobfuscation with cluster-aware graph learning. Our primary innovation leverages LLM’s semantic reconstruction capability to restore syntactic clarity and structural coherence through a multi-stage refinement pipeline, where the language model progressively recovers original code semantics from obfuscated inputs and generates normalized AST representations. By organizing these ASTs into hierarchical graph structures, we deploy a graph transformer with node-to-cluster attention that simultaneously captures: (a) semantically consistent node-level features, and (b) cluster-induced structural patterns, effectively modeling both lexical syntax and program-wide dependency relationships unique to JavaScript’s execution context.

Our key contributions include: 
\begin{itemize}

\item Our approach achieves 94.64\% and 97.71\% F1-scores on the benchmark datasets, with absolute performance gains of 10.74\% and 13.85\% over existing methods. For security-critical low-FPR scenarios (0.0001, 0.001, 0.01), the method demonstrates substantial TPR improvements of 4.82×, 5.91×, and 2.53× relative to the strongest baseline.
\item Our multi-stage LLM deobfuscation pipeline, guided by structured prompt engineering (including string decoding, semantic variable renaming, dynamic invocation reconstruction, and control flow simplification), empirically demonstrates significant deobfuscation improvements. The systematic prompt design ensures: (1) complete payload unpacking, (2) behavioral equivalence preservation, and (3) explanatory metadata generation.
\item We uses a robust dual-scale graph learning framework for JavaScript ASTs that simultaneously models node-level features and cluster-induced structural patterns through node-to-cluster attention, effectively addressing: (1) semantic equivalence dispersion in feature space, and (2) scope chain dependency breakdowns in deep closure nesting.

\end{itemize}

\section{Related Work}
As JavaScript dominates web development, malicious script detection remains an essential security challenge. This work provides a methodological taxonomy of learning-based detectors, analyzing representative approaches across different modeling paradigms while positioning our framework's structural innovations in context.

\subsection{Sequence Model}
Sequence-based models employ dominant sequential learning algorithms (LSTMs, CNNs, and Transformers) to extract patterns from code sequences. While all process code as linear sequences, their inductive biases lead to fundamentally different feature representations.

Recurrent Neural Networks, particularly LSTM \cite{zaremba2014recurrent}, have been adopted for modeling sequential dependencies in malicious JavaScript detection. Fang et al. \cite{fang2018research} used static analysis by learning opcode sequences from compiled JavaScript bytecode using LSTM networks. Subsequent enhancements by Song et al. \cite{song2020malicious} integrated bidirectional LSTMs with program dependence graphs to analyze semantically sliced execution paths, improving resilience against basic obfuscation. Further improvements incorporated attention mechanisms and semantic embeddings \cite{fang2020detecting} to identify critical code segments in tokenized JavaScript. However, these approaches fundamentally suffer from three limitations: (1) linear processing constraints that prevent hierarchical relationship modeling, (2) limited context windows for long-range dependency analysis, and (3) inability to effectively represent non-sequential program structures. 

\begin{sloppypar}
CNNs have been extensively applied to malicious  JavaScript detection through various code representations, including syntax trees, bytecode, and token sequences. The JSAC framework \cite{liang2019jsac} employs parallel CNNs to process both abstract syntax trees (ASTs) and control flow graphs (CFGs), capturing complementary syntactic and semantic features. Rozi et al. \cite{rozi2020deep} advanced this paradigm by introducing a deep pyramid CNN architecture operating on V8 engine bytecode sequences, augmented with recurrent layers for improved obfuscation resilience. Alternative implementations include Sheneamer's stacked CNN ensemble \cite{sheneamer2024vulnerable} for vulnerability detection and ScriptNet's hierarchical CNN \cite{stokes2019scriptnet} for byte-level sequence analysis. While CNNs demonstrate exceptional proficiency in extracting local code patterns and achieving strong static analysis performance, their fundamental architectural constraints—particularly limited receptive fields and absence of explicit structural modeling—severely impair their capability to analyze long-range dependencies or complex program semantics.

Transformer architectures, particularly BERT-based models, have demonstrated promising results in malicious JavaScript detection through contextual token embeddings. While hybrid approaches like the BERT-BiLSTM model \cite{abadeer2022dynamic} improve semantic understanding, they fundamentally lack mechanisms to capture the localized syntactic patterns and hierarchical structural relationships critical for effective malware detection beyond pure semantic analysis.
\end{sloppypar}

\begin{figure*}[t]
    \centering
    \includegraphics[width=\linewidth]{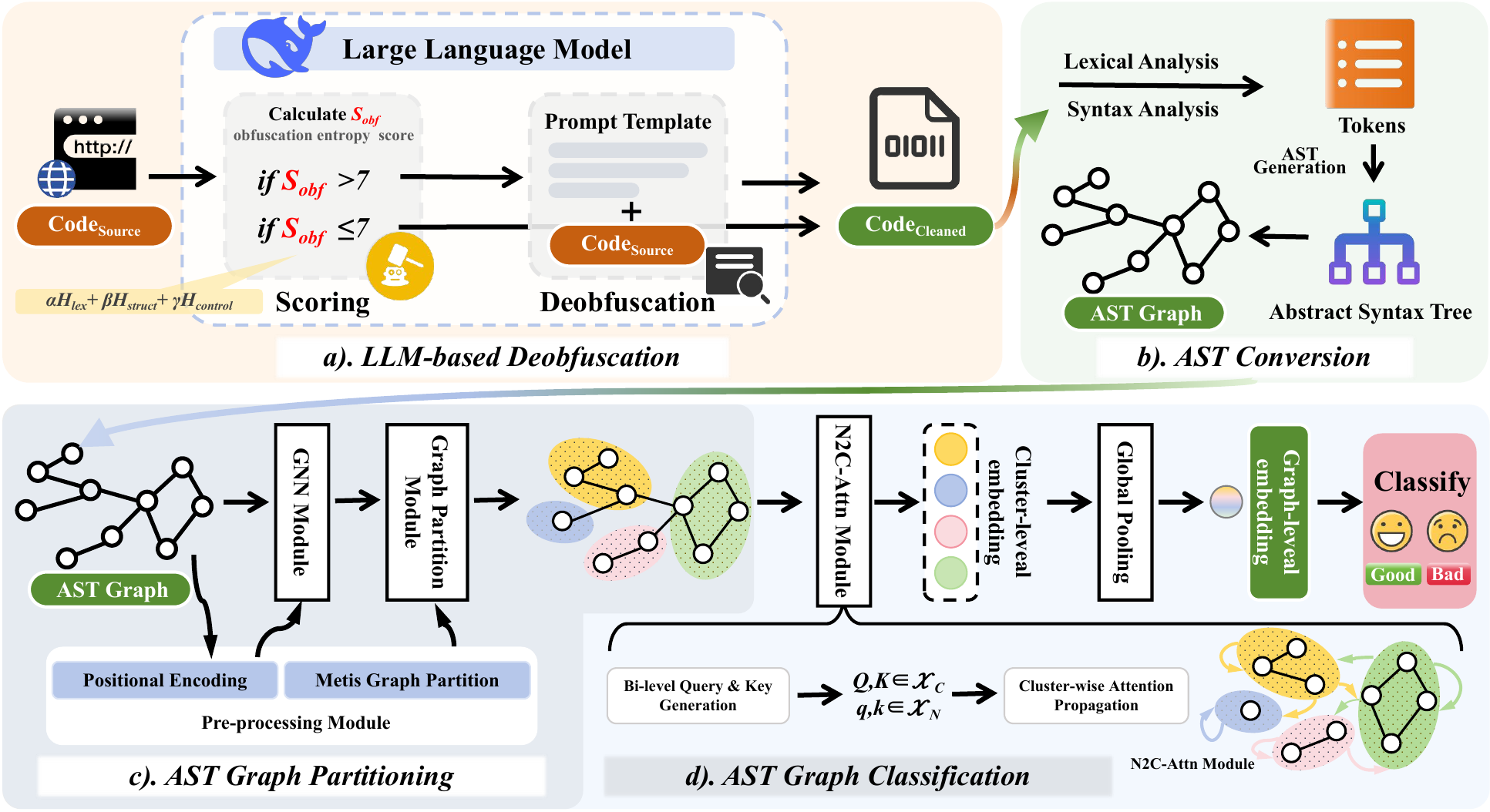}
    \caption{Architecture of the our method.}
    \label{method}
\end{figure*}

\subsection{Code Graph learning}
Source code's intrinsic hierarchical graph structure - encompassing control flows, data dependencies, and lexical scopes - renders GNNs particularly suitable for learning code garph. As demonstrated by \cite{sheng2025dynamic,LIU2025102748,sun2025ethereum}, GNNs excel at modeling complex structural patterns in graph data.  JStrong \cite{fang2022jstrong} apply graph convolutional networks to learn joint structural-semantic features, and JStrack \cite{rozi2021jstrack} employs hierarchical GNNs to preserve syntactic information while analyzing nested code structures.

While GNN applications in code analysis are well-established, their adoption for malicious JavaScript detection remains limited. Existing approaches often employ generic graph architectures (e.g., flat message passing or coarse pooling) that neglect JavaScript-specific characteristics, potentially losing critical syntactic details or facing scalability issues. Our method addresses these limitations through: (1) AST partitioning into meaningful substructures, and (2) node-to-cluster attention for joint hierarchical reasoning, enabling multi-granularity feature integration while preserving structural fidelity. This design demonstrates superior resilience and practical scalability compared to conventional GNN implementations.

\section{Method}

Our proposed method for detecting malicious obfuscated JavaScript code integrates LLM-based deobfuscation with advanced GNN classification. The overall framework is designed to be modular and extensible, allowing it to adapt effectively to diverse obfuscation patterns and attack scenarios. LLMs are leveraged to restore the syntactic and semantic clarity of obfuscated code, after which structural representations such as ASTs are extracted and transformed into graph form. A graph-based classifier, built upon a cluster-aware transformer architecture, is subsequently applied to capture the structural relationships among nodes and clusters, enabling robust and accurate malicious code detection. An overview of the entire detection pipeline is illustrated in Fig.~\ref{method}.

\subsection{LLM-based Deobfuscation}
We utilize DeepSeek-R1, a cost-efficient state-of-the-art large language model, for automated JavaScript deobfuscation. The process incorporates three key components:

\subsubsection{Semantic Representation and Feature Extraction}
The LLM maps obfuscated code $X_{\text{obf}}$ to a semantic space through its pre-trained Transformer architecture:
\begin{equation}
h = \text{Transformer}(X_{\text{obf}}; \theta)
\end{equation}
where $h$ represents the semantic representation vector, and $\theta$ denotes the model parameters. DeepSeek-R1\cite{guo2025deepseek}'s Multi-head Latent Attention (MLA) mechanism extracts key features:
\begin{equation}
c_{KV_t} = W_{D_{KV}} h_t, \quad k_t^C = W_{U_K} c_{KV_t}, \quad v_t^C = W_{U_V} c_{KV_t}
\end{equation}
This allows the model to capture deep semantic relationships within obfuscated code patterns.

\subsubsection{Obfuscation Entropy Scoring}
Before applying deobfuscation, we evaluate the complexity of obfuscation using an entropy-based scoring mechanism. The obfuscation entropy score $S_{\text{obf}}$ is calculated as:
\begin{equation}
S_{\text{obf}} = \alpha H_{\text{lex}} + \beta H_{\text{struct}} + \gamma H_{\text{control}}
\end{equation}
where:
\begin{itemize}
    \item $H_{\text{lex}} = -\sum_{i=1}^{|V|} p_i \log p_i$ represents lexical entropy based on token distribution
    \item $H_{\text{struct}} = \frac{1}{|N|} \sum_{n \in \text{AST}} \log(\text{depth}(n) \cdot \text{children}(n))$ measures structural complexity
    \item $H_{\text{control}} = \frac{|\text{branches}| + |\text{loops}|}{|\text{statements}|}$ quantifies control flow complexity
    \item $\alpha = 0.4$, $\beta = 0.4$, $\gamma = 0.2$ are weighting parameters
\end{itemize}

Only code samples with $S_{\text{obf}} > 7$ undergo deobfuscation, optimizing computational resources while ensuring that genuinely obfuscated code is processed.

\subsubsection{Prompt Engineering for Deobfuscation}
We design a specialized prompt template that guides the LLM to systematically deobfuscate JavaScript code:

\begin{codeprompt}{JavaScript Deobfuscation Prompt Template}
\textbf{Task Description:}\\
Analyze the following obfuscated JavaScript code and provide a clean, readable version while preserving exact functionality.\\[8pt]

\textbf{Instructions:}
\begin{enumerate}
    \item Decode all string obfuscations (hex, base64, unicode escapes)
    \item Replace meaningless variable names with descriptive ones
    \item Unpack compressed/encoded payloads (e.g., eval expressions)
    \item Simplify control flow (remove dead code, flatten conditionals)
    \item Reconstruct function calls from dynamic invocations
    \item Preserve original logic and behavior exactly
\end{enumerate}

\vspace{8pt}
\textcolor{gray}{\hrule height 0.5pt}
\vspace{8pt}

\textbf{Input:} Obfuscated Code\\[4pt]
\texttt{\textcolor{promptblue}{\{obfuscated\_code\}}}\\[8pt]
\textbf{Expected Output:}\\
Provide the deobfuscated version with explanatory comments that describe:
\begin{itemize}
    \item The original obfuscation techniques detected
    \item The deobfuscation steps applied
    \item The restored functionality and control flow
\end{itemize}
\end{codeprompt}

\subsubsection{Probabilistic Generation and Optimization}
The deobfuscation process is modeled as a sequence-to-sequence transformation, where the model predicts the probability distribution of clean code $Y_{\text{clean}}$:
\begin{equation}
P(Y_{\text{clean}} | X_{\text{obf}}) = \prod_{t=1}^T P(y_t | y_{<t}, X_{\text{obf}}; \theta)
\end{equation}

DeepSeek-R1's Mixture of Experts (MoE) architecture enhances generation capability:
\begin{equation}
h'_t = u_t + \sum_{i=1}^{N_s} \text{FFN}^{(s)}_i (u_t) + \sum_{i=1}^{N_r} g_{i,t} \text{FFN}^{(r)}_i (u_t)
\end{equation}
where $g_{i,t}$ is the gating function selecting appropriate experts for deobfuscation patterns.

\subsubsection{Training and Loss Function}
The model is optimized using a deobfuscation-specific loss function:
\begin{equation}
L_{\text{deobf}} = -\sum_{i=1}^N \log P(Y_{\text{clean},i} | X_{\text{obf},i}; \theta) + \lambda L_{\text{semantic}}
\end{equation}
where $L_{\text{semantic}}$ ensures semantic preservation between obfuscated and deobfuscated versions.

\subsubsection{Code Structure Restoration}
The model restores obfuscated structures through syntax tree parsing and semantic inference:
\begin{equation}
\text{AST}_{\text{clean}} = f_{\text{decode}}(h; \theta_{\text{dec}})
\end{equation}
where $\text{AST}_{\text{clean}}$ is the restored abstract syntax tree, and $f_{\text{decode}}$ is the decoding function leveraging the model's generation capabilities.

The deobfuscation pipeline is automated via scripts that interact with the DeepSeek-R1 API, supporting batch processing with configurable parameters (temperature=0.1, max\_tokens=4096) to ensure deterministic and complete deobfuscation. Quality filtering ensures that only successfully deobfuscated samples with semantic equivalence are retained for downstream analysis.

\subsection{AST Conversion}
The deobfuscated code is parsed into ASTs using the Esprima parser for JavaScript. ASTs serve as a natural bridge between source code and graph neural networks by encoding the syntactic structure of programs in a hierarchical graph format.

Given deobfuscated code $Y_{\text{clean}}$, we construct an AST $\mathcal{T} = (\mathcal{V}, \mathcal{E})$ where $\mathcal{V}$ represents AST nodes (e.g., FunctionDeclaration, VariableDeclarator) and $\mathcal{E}$ represents parent-child relationships. The AST is then transformed into a graph representation:
\begin{equation}
\mathcal{G} = (\mathcal{V}, \mathcal{E}_{\text{ext}}, \mathbf{X}, \mathbf{A})
\end{equation}
where $\mathcal{E}_{\text{ext}} = \mathcal{E} \cup \mathcal{E}_{\text{data}} \cup \mathcal{E}_{\text{control}}$ includes extended edges for data flow and control flow.

Node features are encoded through multi-component embedding:
\begin{equation}
\mathbf{x}_i = \mathbf{e}_{\text{type}}(t_i) \oplus \mathbf{e}_{\text{value}}(s_i) \oplus \mathbf{e}_{\text{pos}}(l_i, c_i)
\end{equation}
where $\mathbf{e}_{\text{type}}$, $\mathbf{e}_{\text{value}}$, and $\mathbf{e}_{\text{pos}}$ embed node types, values, and positions respectively.

Finally, the AST graph is converted to PyG format:
\begin{equation}
\text{Data} = \{\mathbf{x}, \text{edge\_index}, \text{edge\_attr}, y\}
\end{equation}
This representation preserves the rich structural information inherent in code while enabling direct application of GNN architectures.

\subsection{AST Graph Partitioning}
To handle large-scale AST graphs and enhance model expressiveness, we apply graph partitioning techniques to divide each AST into multiple patches or subgraphs. This partitioning enables multi-granularity analysis while maintaining computational efficiency.

We employ the METIS algorithm to partition the AST graph into $m$ clusters. The cluster assignment is defined by matrix $\mathbf{C} \in \mathbb{R}^{n \times m}$:
\begin{equation}
\mathbf{C}_{nm}^{\text{Metis}} = \begin{cases} 
\frac{1}{|\mathcal{V}_m|} & \text{if the } n\text{-th node is in the } m\text{-th cluster} \\
0 & \text{otherwise}
\end{cases}
\end{equation}
where $|\mathcal{V}_m|$ denotes the number of nodes in cluster $m$, ensuring balanced representation.

The graph coarsening process transforms the original node features and adjacency matrix to cluster-level representations:
\begin{equation}
\mathbf{X}^P = \mathbf{C}^T \mathbf{X}, \quad \mathbf{A}^P = \mathbf{C}^T \mathbf{A} \mathbf{C}
\end{equation}
where $\mathbf{X}^P \in \mathbb{R}^{m \times d}$ is the cluster-level feature matrix and $\mathbf{A}^P \in \mathbb{R}^{m \times m}$ is the cluster-level adjacency matrix.

For our Cluster-GT architecture, we compute cluster-level queries and keys using the assignment matrix:
\begin{equation}
K_j = \mathbf{W}_k' \left( \sum_s \mathbf{C}_{sj} h_s \right), \quad Q_i = \mathbf{W}_v' \left( \sum_s \mathbf{C}_{si} h_s \right)
\end{equation}
This formulation enables the Node-to-Cluster Attention mechanism to integrate both node-level and cluster-level information, capturing hierarchical patterns essential for detecting malicious code structures. The partitioned graphs retain both local syntactic details and global structural patterns, supporting robust classification even in the presence of obfuscation.

\subsection{Cluster-wise Code Graph Transformer}
In JavaScript ASTs, dynamic typing scatters semantically similar nodes while deeply nested functions fracture scope capturing, introducing structural noise and semantic ambiguity.To address these challenges, we propose a cluster-based AST graph learning framework employing Cluster-wise Graph Transformer \cite{huang2024cluster}, that employs node-to-cluster attention(N2C-Attn) and cluster-guided message passing to capture consistent linguistic patterns across node types and bridge semantic relationships fractured by scope boundaries.

\subsubsection{Node-to-Cluster Attention} The general form of the N2C-Attn is defined as:
\begin{equation}
\text{N2C-Attn}(X)_i = \frac{\sum_j \mathbf{A}_{i,j}^P \sum_t \mathbf{C}_{tj} \, \kappa_{\text{B}}(\{Q_i, q_i\}, \{K_j, k_t\}) v_t}{\sum_j \mathbf{A}_{i,j}^P \sum_t \mathbf{C}_{tj} \, \kappa_{\text{B}}(\{Q_i, q_i\}, \{K_j, k_t\})}
\end{equation}
where $\mathbf{A}^P$ is the cluster-level adjacency matrix, $\mathbf{C}$ is the cluster assignment matrix, $\kappa_{\text{B}}$ is the dual-granularity kernel function, and $Q_i, q_i, K_j, k_t$ are the cluster-level and node-level queries and keys, respectively.

For the tensor product kernel N2C-Attn-T:
\begin{equation}
\kappa_{\text{B}}(\{Q_i, q_i\}, \{K_j, k_t\}) = \kappa_C(Q_i, K_j) \, \kappa_N(q_i, k_t)
\end{equation}
\begin{equation}
\text{N2C-Attn-T}(X)_i = \frac{\sum_j \mathbf{A}_{i,j}^P \sum_t \mathbf{C}_{tj} \, \kappa_C(Q_i, K_j) \, \kappa_N(q_i, k_t) v_t}{\sum_j \mathbf{A}_{i,j}^P \sum_t \mathbf{C}_{tj} \, \kappa_C(Q_i, K_j) \, \kappa_N(q_i, k_t)}
\end{equation}

For the convex linear combination kernel (N2C-Attn-L):
\begin{equation}
\kappa_{\text{B}}(\{Q_i, q_i\}, \{K_j, k_t\}) = \alpha \, \kappa_C(Q_i, K_j) + \beta \, \kappa_N(q_i, k_t)
\end{equation}
{\small{
\begin{equation}
\text{N2C-Attn-L}(X)_i = \frac{\sum_j \mathbf{A}_{i,j}^P \sum_t \mathbf{C}_{tj} (\alpha \, \kappa_C(Q_i, K_j) + \beta \, \kappa_N(q_i, k_t)) v_t}{\sum_j \mathbf{A}_{i,j}^P \sum_t \mathbf{C}_{tj} (\alpha \, \kappa_C(Q_i, K_j) + \beta \, \kappa_N(q_i, k_t))}
\end{equation}
}}
where $\alpha, \beta \geq 0$ and $\alpha + \beta = 1$ are learnable parameters.

\subsubsection{Dual-granularity Kernelized Attention Framework} The general form of the dual-granularity kernel function is:
\begin{equation}
\kappa_{\text{B}}(\{\mathbf{x}^m\}_{m=1}^M, \{\mathbf{y}^m\}_{m=1}^M) = f_\eta(\{\kappa_m(\mathbf{x}^m, \mathbf{y}^m)\}_{m=1}^M)
\end{equation}
where $M=2$ in N2C-Attn, corresponding to node-level and cluster-level kernels.

\subsubsection{AST Graph-level Embedding} The final graph-level embedding is obtained by average pooling over the cluster outputs of the N2C-Attn module:
\begin{equation}
\mathbf{h}_{\text{graph}} = \frac{1}{|\mathcal{N}^P|} \sum_{i \in \mathcal{N}^P} \text{N2C-Attn}(X)_i
\end{equation}
where $\mathcal{N}^P$ denotes the set of clusters and $\text{N2C-Attn}(X)_i$ is the attention output for the $i$-th cluster.

\subsubsection{Multi-Layer Perceptron  for Classification} A Multi-Layer Perceptron\cite{riedmiller2014multi} is a feedforward neural network used for classification tasks, consisting of an input layer with \( m \) neurons (features), \( L-1 \) hidden layers with \( n_l \) neurons each, and an output layer with \( K \) neurons (classes).

For each hidden layer \( l \in \{1, \dots, L-1\} \), the computation is:
\begin{align}
    \mathbf{z}^{(l)} &= \mathbf{W}^{(l)} \mathbf{a}^{(l-1)} + \mathbf{b}^{(l)}, \\
    \mathbf{a}^{(l)} &= f(\mathbf{z}^{(l)}),
\end{align}
where \(\mathbf{W}^{(l)} \in \mathbb{R}^{n_l \times n_{l-1}}\) is the weight matrix, \(\mathbf{b}^{(l)} \in \mathbb{R}^{n_l}\) is the bias vector, and \( f \) is an activation function (e.g., ReLU, tanh).

For the output layer (\( l = L \)):
\begin{itemize}
    \item \textbf{Binary Classification}: The output is
    \[
    \hat{y} = \sigma(\mathbf{W}^{(L)} \mathbf{a}^{(L-1)} + \mathbf{b}^{(L)}) = \frac{1}{1 + e^{-z^{(L)}}},
    \]
    where \(\sigma\) is the sigmoid function.
    \item \textbf{Multi-class Classification}: The output is
    \[
    \hat{y}_i = \text{softmax}(\mathbf{z}^{(L)})_i = \frac{e^{z_i^{(L)}}}{\sum_{j=1}^K e^{z_j^{(L)}}}, \quad \text{for } i = 1, \dots, K.
    \]
\end{itemize}

\begin{itemize}
    \item \textbf{Binary Cross-Entropy}: For binary classification, the loss is:
    \[
    \mathcal{L} = -\left[ y \log(\hat{y}) + (1-y) \log(1-\hat{y}) \right].
    \]
    \item \textbf{Categorical Cross-Entropy}: For multi-class classification, the loss is:
    \[
    \mathcal{L} = -\sum_{i=1}^K y_i \log(\hat{y}_i).
    \]
\end{itemize}

\section{Experimental Setup}
\subsection{Dataset}
To evaluate our JavaScript detector, we constructed two new datasets to assess its performance. These datasets are designed to provide a balanced and diverse representation of both benign and malicious JavaScript samples. The detailed class distribution of these datasets is summarized in Table~\ref{tab:dataset_composition}. The datasets are described as follows:

\begin{itemize}
\item \textbf{Dataset$_1$}: This corpus is composed of a large-scale benign dataset, JS150~\cite{js150k}, and a publicly available malicious corpus, JS-Malicious-Dataset~\cite{js_malicious}. A subset of samples from JS150 is included to ensure balanced coverage of both classes while preserving real-world diversity.
\item \textbf{Dataset$_2$}: Sourced from a GitHub repository~\cite{js_dataset}, this dataset separates benign (\texttt{goodjs}) and malicious (\texttt{badjs}) scripts. It includes all 8,079 benign and 8,500 malicious samples.
\end{itemize}

We employ the DeepSeek code model to automatically deobfuscate obfuscated JavaScript code. We utilize carefully crafted prompt templates to detect and decode string encodings (e.g., Hex or Base64), standardize variable names, and reconstruct control flows while preserving the original functionality. The deobfuscated code is then transformed into an Abstract Syntax Tree (AST) with data flow annotations using the Esprima parser. For sequence-based models (e.g., BERT or LSTM), we perform language feature segmentation, while for graph-based models (e.g., GNN), we apply the METIS algorithm to partition the AST into 8 subgraphs compatible with the PyG format. For model development, we randomly partition the combined corpus, allocating 80\% for training, 10\% for validation, and 10\% as a held-out test set, ensuring the benign-to-malicious ratio is maintained across all splits.

For model development we randomly sample 80\% of the combined corpus for training, 10\% for validation, and retain 10\% as a held-out test set whilst preserving the benign/malicious ratio.

\begin{table}[!t]
    \centering
    \caption{Composition of the Dataset$_{1}$ and Dataset$_{2}$}
    \label{tab:dataset_composition}
    \resizebox{1.0\linewidth}{!}{
    \begin{tabular}{lccc}
        \toprule
        \textbf{Source} & \textbf{Benign} & \textbf{Malicious} & \textbf{Total}\\
        \midrule\midrule
        
        JS150 & 150\,000 & 0 & 150\,000\\
        JS-Malicious-Dataset & 0 & 1\,357 & 1\,357\\
        \midrule
        \textbf{Dataset$_{1}$} & \textbf{150\,000} & \textbf{1\,357} & \textbf{151\,357}\\
        \midrule\midrule
        JavaScript\_Datasets & 8\,079 & 8\,500 & 16\,579\\
        
        \midrule
        \textbf{Dataset$_{2}$} & \textbf{8\,079} & \textbf{8\,500} & \textbf{16\,579}\\
        
        \bottomrule
    \end{tabular}}
\end{table}

\subsection{Baselines}
We compare our method against the following baseline models:
\begin{itemize}
    \item \textbf{BERT}: A Transformer-based pre-trained language model that revolutionized natural language processing through bidirectional contextual awareness. Its core architecture employs 12 Transformer encoder layers (768 hidden dimensions), pre-trained on text corpora (e.g., Wikipedia) using Masked Language Modeling (MLM) and Next Sentence Prediction (NSP) objectives. 
    
    \item \textbf{CodeBERT\cite{feng2020codebert}}: A specialized pre-trained language model designed for programming-related tasks, building upon the Transformer architecture. Unlike general-purpose language models like BERT, CodeBERT is uniquely trained on both programming languages  and natural language text, enabling it to understand the nuanced relationship between code and human language. The model employs innovative training objectives including MLM and RTD, which specifically enhance its ability to comprehend code syntax and semantics. 
    
    \item \textbf{LSTM}: A bidirectional LSTM with 2 layers, 256 hidden dimensions, dropout of 0.2, and an embedding dimension of 128. The model is trained using a batch size of 32 and an initial learning rate of 1e-3 with a step-wise decay schedule. 
    
    \item \textbf{CNN}: A 1D convolutional architecture with 3 layers, kernel sizes of [3, 5, 7], 128 filters per layer, ReLU activations, and global max pooling. The model is trained with a batch size of 32 and a learning rate of 5e-4. 
    
    \item \textbf{GCN\cite{kipf2016semi}}: A Graph Convolutional Network with 3 layers, 128 hidden dimensions, ReLU activations, and dropout of 0.5. The model processes AST graphs using message passing between adjacent nodes, with graph-level pooling for final classification. We train with a batch size of 16 and a learning rate of 5e-4 with weight decay of 1e-5. 
\end{itemize}

All baseline models process tokenized code sequences and are trained using the Adam optimizer with weight decay of 1e-5. Early stopping with a patience of 10 epochs based on validation F1-score is employed across all models.

\subsection{Evaluation Metrics}
We evaluate model performance using the following metrics:

\begin{itemize}
    \item \textbf{Accuracy}: The proportion of correctly classified samples.
    \begin{equation}
    \text{Accuracy} = \frac{TP + TN}{TP + TN + FP + FN}
    \end{equation}
    
    \item \textbf{Precision}: The proportion of true positive predictions among all positive predictions.
    \begin{equation}
    \text{Precision} = \frac{TP}{TP + FP}
    \end{equation}
    
    \item \textbf{Recall}: The proportion of true positive predictions among all actual positives.
    \begin{equation}
    \text{Recall} = \frac{TP}{TP + FN}
    \end{equation}
    
    \item \textbf{F1-score}: The harmonic mean of precision and recall.
    \begin{equation}
    \text{F1-score} = 2 \times \frac{\text{Precision} \times \text{Recall}}{\text{Precision} + \text{Recall}}
    \end{equation}
    
    \item \textbf{AUC-ROC}: The area under the Receiver Operating Characteristic curve, which plots the true positive rate against the false positive rate at various threshold settings.
    \begin{equation}
    \text{AUC-ROC} = \int_{0}^{1} TPR(FPR^{-1}(t)) \, dt
    \end{equation}
    where $TPR$ is the true positive rate and $FPR$ is the false positive rate.
\end{itemize}

For malicious code detection, we consider the malicious class as the positive class. These metrics provide a comprehensive evaluation of model performance, with F1-score being particularly important due to the security implications of both false positives and false negatives in malware detection.

\subsection{Implementation Details}
All experiments were conducted using PyTorch 1.12.0 and PyTorch Geometric 2.2.0 on an NVIDIA RTX 3090 GPU with 24GB RAM.    Our proposed Cluster-GT model uses 3 Cluster-GT layers with 128-dimensional hidden states, 8 attention heads per layer, and dropout rate of 0.1.    We employed the METIS algorithm to partition each AST into 8 clusters depending on graph size.  Hyperparameters for all models were optimized via grid search on the validation set using a 10-fold cross-validation strategy.     For the dual-granularity kernelized attention mechanism, we used the tensor product kernel (N2C-Attn-T). The DeepSeek-Coder-7B-Instruct\cite{guo2024deepseekcoderlargelanguagemodel} model was accessed through an API interface for the deobfuscation pipeline.

\section{Results and Discussion}

\subsection{Overall Performance Comparison}
To evaluate the comprehensive performance of our proposed method, we conducted a comparative analysis against a suite of baseline models on Dataset$_{1}$ and Dataset$_{2}$, with the results presented in Table~\ref{tab:overall_comparison}. In this evaluation, our method leverages its integrated LLM-based deobfuscation pipeline, whereas all baseline models (BERT, CodeBERT, LSTM, CNN, and GCN) were evaluated on the original, potentially obfuscated datasets. This experimental design aims to simulate a realistic scenario, pitting our end-to-end solution against existing standard methodologies.

The results clearly demonstrate the decisive superiority of our method. On Dataset$_{1}$, our method achieves an F1-score of 0.9464 and an AUC of 0.9750, significantly outperforming the best baseline, CodeBERT, which recorded an F1-score of 0.8551 and an AUC of 0.9242. This equates to a relative improvement of approximately 9.13\% in F1-score and 5.08\% in AUC. The performance gap widened on the more challenging Dataset$_{2}$, where our method attained an F1-score of 0.9771 and an AUC of 0.9818. In contrast, CodeBERT achieved an F1-score of 0.8605 and an AUC of 0.9430, marking a relative improvement for our method of approximately 11.66\% in F1-score and 3.88\% in AUC.

The baseline models exhibit pronounced limitations when processing unprocessed data. Sequence-based models like BERT and LSTM are particularly vulnerable, as code obfuscation techniques such as variable renaming and control flow flattening disrupt the sequential semantics upon which they depend. Even the graph-based GCN model, despite utilizing Abstract Syntax Trees (ASTs), only achieved an F1-score of 0.8674 on Dataset$_{2}$, indicating its inability to capture the multi-granularity structural information that our node-to-cluster attention mechanism effectively extracts.

\begin{table*}[tbp]
    \centering
    \caption{Performance Comparison of Our Method Against Baselines on Dataset$_{1}$ and Dataset$_{2}$. Note: Values in parentheses () indicate the performance gap between each baseline method and our proposed approach}
    \label{tab:overall_comparison}
    \fontsize{14}{14}\selectfont 
    \renewcommand{\arraystretch}{1.2} 
    
    \resizebox{1.0\linewidth}{!}{
    \begin{tabular}{lcccc cccc}
    \toprule
    \multirow{2}{*}{\textbf{\large Model}} & \multicolumn{4}{c}{\textbf{\large Dataset$_{1}$}} & \multicolumn{4}{c}{\textbf{\large Dataset$_{2}$}} \\
    \cmidrule(lr){2-5} \cmidrule(lr){6-9}
    & \textbf{Precision} & \textbf{Recall} & \textbf{F1-Score} & \textbf{AUC} & \textbf{Precision} & \textbf{Recall} & \textbf{F1-Score} & \textbf{AUC} \\
    \midrule
    BERT & 0.8216\,(-13.10) & 0.8571\,(-8.32) & 0.8390\,(-10.74) & 0.9150\,(-6.00) & 0.8316\,(-13.95) & 0.8771\,(-10.61) & 0.8537\,(-12.34) & 0.9242\,(-5.76) \\[5pt]
    LSTM & 0.7552\,(-19.74) & 0.8650\,(-7.53) & 0.8064\,(-14.00) & 0.8700\,(-10.50) & 0.8000\,(-17.11) & 0.8811\,(-10.21) & 0.8386\,(-13.85) & 0.8911\,(-9.07) \\[5pt]
    CodeBERT & 0.8489\,(-10.37) & 0.8615\,(-7.88) & 0.8551\,(-9.13) & 0.9242\,(-5.08) & 0.8701\,(-10.10) & 0.8511\,(-13.21) & 0.8605\,(-11.66) & 0.9430\,(-3.88) \\[5pt]
    GCN & 0.8325\,(-12.01) & 0.8623\,(-7.80) & 0.8471\,(-9.93) & 0.9189\,(-5.61) & 0.8737\,(-9.74) & 0.8611\,(-12.21) & 0.8674\,(-10.97) & 0.9480\,(-3.38) \\[5pt]
    CNN & 0.8071\,(-14.55) & 0.8485\,(-9.18) & 0.8273\,(-11.91) & 0.8350\,(-14.00) & 0.8471\,(-12.40) & 0.8585\,(-12.47) & 0.8528\,(-12.43) & 0.9150\,(-6.68) \\[5pt]
    \midrule
    \textbf{\large Our Method} & \textbf{0.9526} & \textbf{0.9403} & \textbf{0.9464} & \textbf{0.9750} & \textbf{0.9711} & \textbf{0.9832} & \textbf{0.9771} & \textbf{0.9818} \\
    \bottomrule
    \end{tabular}}
\end{table*}

\begin{table*}[htbp]
    \centering
    \caption{Performance Impact of LLM-based Deobfuscation Across All Models}
    \label{tab:deobfuscation_impact}
    \resizebox{1.0\linewidth}{!}{
    \begin{tabular}{llcccc cccc}
    \toprule
    \multirow{2}{*}{\textbf{Model}} & \multirow{2}{*}{\textbf{Condition}} & \multicolumn{4}{c}{\textbf{Dataset$_{1}$}} & \multicolumn{4}{c}{\textbf{Dataset$_{2}$}} \\
    \cmidrule(lr){3-6} \cmidrule(lr){7-10}
    & & Precision & Recall & F1-Score & AUC & Precision & Recall & F1-Score & AUC \\
    \midrule
    \multirow{3}{*}{BERT} & Obfuscated & 0.8216 & 0.8571 & 0.8390 & 0.9150 & 0.8316 & 0.8771 & 0.8537 & 0.9242 \\
    & Deobfuscated & 0.8914 & 0.8732 & 0.8822 & 0.9112 & 0.9014 & 0.8832& 0.8922& 0.9400 \\
    \cmidrule(lr){2-10}
    & \textit{Improvement (\%)} & +6.98 & +1.61 & +4.32 & -0.38 & +6.98 & +0.61 & +3.85 & +1.58 \\
    \midrule
    \multirow{3}{*}{LSTM} & Obfuscated & 0.7552 & 0.8650 & 0.8064 & 0.8700 & 0.8000 & 0.8811 & 0.8386 & 0.8911 \\
    & Deobfuscated & 0.8026 & 0.9212& 0.8578 & 0.9090 & 0.8516 & 0.9351 & 0.8914& 0.9332 \\
    \cmidrule(lr){2-10}
    & \textit{Improvement (\%)} & +4.74 & +5.62 & +5.14 & +3.90 & +5.16 & +5.40 & +5.28 & +4.21 \\
    \midrule
    \multirow{3}{*}{CodeBERT} & Obfuscated & 0.8489 & 0.8615 & 0.8551 & 0.9242& 0.8701 & 0.8511 & 0.8605 & 0.9430 \\
    & Deobfuscated & 0.9087 & 0.8882 & 0.8983 & 0.9422 & 0.9172 & 0.8963& 0.9066& 0.9601 \\
    \cmidrule(lr){2-10}
    & \textit{Improvement (\%)} & +5.98 & +2.67 & +4.32 & +1.80 & +4.71 & +4.52 & +4.61 & +1.71 \\
    \midrule
    \multirow{3}{*}{GCN} & Obfuscated & 0.8325 & 0.8623 & 0.8471 & 0.9189 & 0.8737 & 0.8611 & 0.8674 & 0.9480 \\
    & Deobfuscated & 0.9323 & 0.8611 & 0.8952 & 0.9512 & 0.9223 & 0.8582& 0.8885& 0.9611 \\
    \cmidrule(lr){2-10}
    & \textit{Improvement (\%)} & +9.98 & -0.12 & +4.81 & +3.23 & +4.86 & -0.29 & +2.11 & +1.31 \\
    \midrule
    \multirow{3}{*}{CNN} & Obfuscated & 0.8071 & 0.8485 & 0.8273 & 0.8350 & 0.8471 & 0.8585 & 0.8528 & 0.9150 \\
    & Deobfuscated & 0.8836 & 0.8936 & 0.8886& 0.9350 & 0.9036& 0.8936 & 0.8986 & 0.9450 \\
    \cmidrule(lr){2-10}
    & \textit{Improvement (\%)} & +7.65 & +4.51 & +6.13 & +10.00 & +5.65 & +3.51 & +4.58 & +3.00 \\
    \midrule
    \multirow{3}{*}{\textbf{Our Method}} & Obfuscated & 0.9100 & 0.9000 & 0.9050 & 0.9423 & 0.9201 & 0.9311 & 0.9256 & 0.9542 \\
    & \textbf{Deobfuscated} & \textbf{0.9526} & \textbf{0.9403} & \textbf{0.9464} & \textbf{0.9750} & \textbf{0.9711} & \textbf{0.9832} & \textbf{0.9771} & \textbf{0.9818} \\
    \cmidrule(lr){2-10}
    & \textit{\textbf{Improvement (\%)}} & \textbf{+4.26} & \textbf{+4.03} & \textbf{+4.14} & \textbf{+3.27} & \textbf{+5.10} & \textbf{+5.21} & \textbf{+5.15} & \textbf{+2.76} \\
    \bottomrule
    \end{tabular}}
\end{table*}

\begin{figure}
    \centering
    \includegraphics[width=0.95\linewidth]{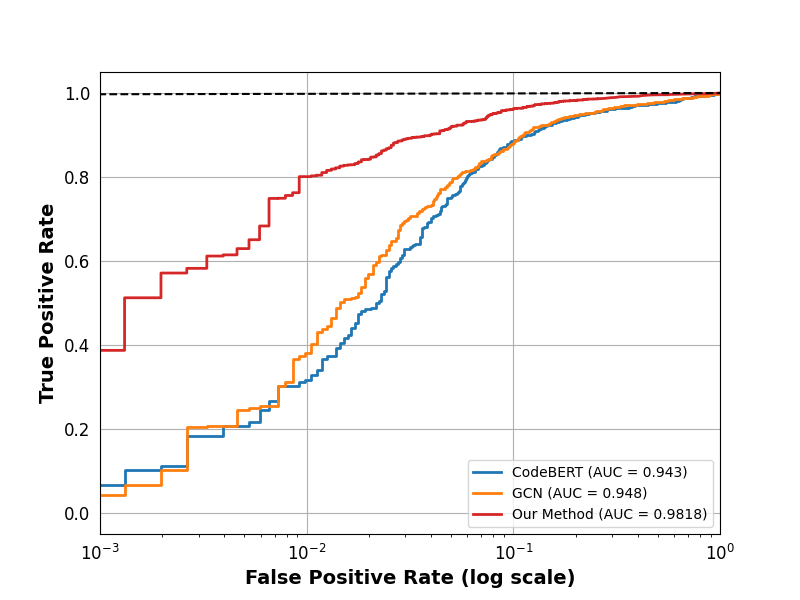}
    \caption{ROC curves}
    \label{fig:roc}
\end{figure}

To further investigate the models’ behavior under strict false positive constraints, we evaluated the True Positive Rate (TPR) at various low False Positive Rate (FPR) levels. As shown in Table~\ref{tab:model_performance}, our method consistently outperforms both CodeBERT and GCN across all FPR thresholds, particularly excelling in extremely low-FPR regimes (e.g., 0.2473 TPR at 0.0001 FPR). The corresponding ROC curves in Figure~\ref{fig:roc} further illustrate this advantage, with our method achieving the highest AUC of 0.9818 and dominating other baselines across the entire FPR range, especially in the low-FPR region relevant to security-critical scenarios.

The exceptional performance of our method is attributable to its core design principles:
\begin{itemize}
    \item \textbf{Integrated Deobfuscation Pipeline:} By normalizing code representations via an LLM prior to feature extraction, our method fundamentally reduces the complexity of the detection task.
    \item \textbf{Robust Graph-based Representation:} The AST-based graph structure preserves critical syntactic and logical relationships, offering resilience against structural obfuscation that sequence-based models lack.
    \item \textbf{Multi-granularity Feature Fusion:} The node-to-cluster attention mechanism enables the model to capture features at multiple levels of abstraction, from local code constructs to global patterns, facilitating a deeper understanding of the code's behavior.
\end{itemize}
These advantages establish our method as a superior solution in terms of precision and generalization, making it an ideal choice for demanding security applications.

\begin{table}[htbp]
    \centering
    \caption{Controlled FPR Evaluation (TPR @ FPR Levels) on Dataset$_{2}$. Note: Values in parentheses indicate the performance multiplier of our method versus each baseline}
    \label{tab:model_performance}
    \resizebox{1.0\linewidth}{!}{
    \begin{tabular}{lcccc}
    \toprule
    \textbf{Model} & \multicolumn{4}{c}{\textbf{TPR @ FPR Level}} \\
    \cmidrule(lr){2-5}
    & 0.0001 & 0.001 & 0.01 & 0.1 \\
    \midrule
    \multirow{2}{*}{CodeBERT} & 0.0513 & 0.0655 & 0.3171 & 0.8864 \\
    & (4.82$\times$) & (5.91$\times$) & (2.53$\times$) & (1.09$\times$) \\
    \midrule  
    \multirow{2}{*}{GCN} & 0.0189 & 0.0425 & 0.3813 & 0.8816 \\
    & (13.09$\times$) & (9.11$\times$) & (2.10$\times$) & (1.09$\times$) \\
    \midrule
    \textbf{Our Method} & \textbf{0.2473} & \textbf{0.3872} & \textbf{0.8008} & \textbf{0.9624} \\
    \bottomrule
    \end{tabular}}
\end{table}

\subsection{Impact of Deobfuscation}
Table~\ref{tab:deobfuscation_impact} illustrate the performance of models on Dataset$_1$ and Dataset$_2$ before and after applying our LLM-based deobfuscation pipeline, revealing substantial improvements that highlight the effectiveness of this approach in normalizing code structures and enhancing feature extraction.

On Dataset$_1$, BERT’s F1-score rises from 0.8390 in obfuscated code to 0.8822 after deobfuscation, a 4.32\% improvement, while LSTM shows a more pronounced gain, increasing from 0.8064 to 0.8578, or 5.14\%. CodeBERT improves from 0.8551 to 0.8983, a 4.32\% boost, and CNN advances from 0.8273 to 0.8886, achieving a 6.13\% enhancement. On Dataset$_2$, similar trends emerge: BERT’s F1-score grows from 0.8537 to 0.8922, a 3.85\% increase; LSTM improves from 0.8386 to 0.8914, a 5.28\% gain; CodeBERT rises from 0.8605 to 0.9066, up 4.61\%; and CNN increases from 0.8528 to 0.8986, a 4.58\% improvement.

Graph-based models also benefit, though to a lesser extent. GCN’s F1-score on Dataset$_1$ improves from 0.8471 to 0.8952, a 4.81\% increase, and on Dataset$_2$ from 0.8674 to 0.8885, a 2.11\% gain. Our method, already robust, sees its F1-score rise from 0.9050 to 0.9464 on Dataset$_1$, a 4.14\% improvement, and from 0.9256 to 0.9771 on Dataset$_2$, a 5.15\% enhancement.

Sequence-based models collectively achieve an average F1-score improvement of 4.98\% on Dataset$_1$ and 4.58\% on Dataset$_2$, significantly higher than the 4.48\% and 3.63\% averages for graph-based models on the respective datasets. This disparity underscores the greater sensitivity of sequence-based models to obfuscation techniques, such as variable renaming and control flow manipulation, which disrupt sequential token relationships. Our method, leveraging robust graph-based representations and multi-granularity feature integration, maintains superior performance in both obfuscated and deobfuscated scenarios, with minimal performance gaps. The consistent enhancements across all models affirm the critical role of our LLM-based deobfuscation pipeline in boosting detection accuracy, particularly for sequence-based models, while our method’s inherent resilience ensures unmatched reliability.

\begin{table*}[t]
    \centering
    \caption{Performance of Our Method Across Different Node Counts and Cluster Kernels}
    \label{tab:scalability_study}
    \resizebox{1.0\linewidth}{!}{
    \begin{tabular}{l ccc ccc ccc}
        \toprule
        \multirow{2}{*}{\textbf{Cluster Kernels}} & \multicolumn{3}{c}{\textbf{1000 Nodes}} & \multicolumn{3}{c}{\textbf{2000 Nodes}} & \multicolumn{3}{c}{\textbf{4000 Nodes}} \\
        \cmidrule(lr){2-4} \cmidrule(lr){5-7} \cmidrule(lr){8-10}
        & Accuracy & Recall & F1-Score & Accuracy & Recall & F1-Score & Accuracy & Recall & F1-Score \\
        \midrule
        1 & 0.9346 & 0.9281 & 0.9314 & 0.9346 & 0.9192 & 0.9268 & 0.9542 & 0.9242 & 0.9379 \\
        4 & 0.9474 & 0.9374 & 0.9425 & 0.9673 & 0.9573 & 0.9620 & 0.9671 & 0.9671 & 0.9670 \\
        8 & 0.9671 & 0.9539 & 0.9607 & 0.9769 & 0.9605 & 0.9686 & 0.9868 & 0.9871 & 0.9870 \\
        \bottomrule
    \end{tabular}}

\end{table*}

\subsection{Analysis of Architectural Advantage}
Table~\ref{tab:deobfuscation_impact} demonstrates that our method's superior performance is not solely dependent on its deobfuscation pipeline but is also rooted in its advanced architectural design.

When all models were evaluated on the original obfuscated datasets, our method achieved F1-scores of 0.9050 and 0.9256 on Dataset$_1$ and Dataset$_2$, respectively. These results significantly surpass those of the best-performing baseline, CodeBERT, which scored 0.8551 and 0.8605. This indicates that our model's graph-based structure is inherently more resilient to common obfuscation techniques. Furthermore, this performance gap persists on the deobfuscated datasets. Our method obtains F1-scores of 0.9464 and 0.9771, while CodeBERT reaches 0.8983 and 0.9066, confirming our model's superior capability to learn from normalized code representations.

This consistent outperformance is attributable to several core architectural features:
\begin{itemize}
    \item \textbf{Preservation of hierarchical code structure:} By leveraging ASTs, our model effectively preserves the code's hierarchical integrity, enabling a more robust detection of malicious patterns that are agnostic to superficial changes in code syntax.
    \item \textbf{Multi-granularity feature integration:} The Node-to-Cluster Attention mechanism facilitates the integration of features at various abstraction levels, allowing the model to capture both fine-grained local patterns and coarse-grained global patterns.
    \item \textbf{Superior precision and generalization:} The architecture demonstrates excellent generalization, with the F1-score difference between Dataset$_1$ and Dataset$_2$ for our final model being minimal. This showcases its ability to maintain high performance on diverse codebases.
\end{itemize}
These design choices collectively create a more powerful and reliable model for malicious code detection, independent of the benefits provided by the deobfuscation pre-processing.

\subsection{Scalability Analysis}

Table~\ref{tab:scalability_study} presents the performance of our method across varying numbers of nodes (1000, 2000, 4000) and cluster kernels (1, 4, 8) in the Abstract Syntax Tree (AST) graphs, evaluated using accuracy, recall, and F1-score. These results demonstrate the scalability of our method, highlighting its ability to effectively integrate structural information as graph size increases and its enhanced performance with finer cluster granularity.

As the number of nodes in the AST graph increases from 1000 to 4000, our method consistently improves across all metrics. For instance, with eight cluster kernels, the F1-score rises from 0.9607 at 1000 nodes to 0.9870 at 4000 nodes, a 2.7\% improvement. This trend underscores the method’s ability to leverage richer structural information in larger graphs, enabling more precise detection of malicious code patterns. Similarly, accuracy and recall exhibit significant gains, reaching 0.9868 and 0.9871, respectively, at 4000 nodes with eight clusters, reflecting a balanced performance that minimizes both false positives and false negatives.

The impact of cluster granularity is particularly pronounced in larger graphs. At 4000 nodes, increasing the number of cluster kernels from one to eight boosts the F1-score from 0.9379 to 0.9870, a 5.2\% enhancement, compared to a 3.1\% improvement (0.9314 to 0.9607) at 1000 nodes. This indicates that finer cluster divisions are more effective in large-scale graphs, where diverse structural patterns require nuanced feature integration. In contrast, with a single cluster kernel, performance remains suboptimal across all node counts, with an F1-score of only 0.9379 at 4000 nodes, highlighting the critical role of the Node-to-Cluster Attention mechanism in capturing multi-granularity features.

Furthermore, the performance gains from increasing cluster kernels diminish slightly at smaller node counts. For example, at 1000 nodes, the F1-score improves by 1.9\% from four to eight clusters (0.9425 to 0.9607), compared to 2.1\% at 4000 nodes (0.9670 to 0.9870). This suggests that our method’s scalability is most pronounced in larger graphs, where additional clusters yield greater benefits. The balanced performance across metrics at 4000 nodes and eight clusters, with near-identical accuracy, recall, and F1-score, further demonstrates the method’s robustness in handling complex graph structures.

These findings confirm that our method scales effectively with increasing graph size and cluster granularity, making it well-suited for real-world applications where AST graphs may vary significantly in scale and complexity.

\section{Limitations and Future Work}
The experimental results demonstrate that our method achieves significant performance improvements over the baseline approaches. In this section, we analyze the limitations of the proposed method and discuss potential opportunities for future enhancements:

\begin{itemize}
    \item \textbf{Limitations of AST-Based Representation}: Our approach primarily leverages Abstract Syntax Trees (ASTs) for code representation. While ASTs capture rich structural and syntactic information, alternative representations such as Control Flow Graphs (CFGs), Program Dependency Graphs (PDGs), and bytecode/IR-based features can provide complementary semantic and behavioral insights. Future work should explore hybrid representations combining ASTs with these techniques to enhance detection accuracy and robustness.
    \item \textbf{Limitations in Dynamic Feature Analysis}: While our method incorporates two JavaScript-specific characteristics, its handling of JavaScript's dynamic nature remains incomplete. Unlike other programming languages where malicious behavior is often statically identifiable, JavaScript's malicious patterns frequently manifest during dynamic execution (e.g., through eval(), dynamic property access, or runtime code generation). Future work could integrate dynamic analysis through browser instrumentation or execution tracing to capture runtime behaviors, subsequently encoding these features as additional node/edge attributes in our graph representation.
    \item \textbf{Limitations in Model Selection}: Our approach primarily employs DeepSeek's base version for deobfuscation tasks, with our core contribution lying in the refined prompt engineering for large language models. While utilizing more advanced LLM versions (such as DeepSeek-V2 or GPT-4) could potentially yield better deobfuscation results, we deliberately focused on prompt optimization and were constrained by the prohibitive costs of premium LLM APIs. Future work should systematically evaluate the cost-benefit tradeoffs of state-of-the-art LLMs across different JavaScript obfuscation patterns, particularly for dynamic features like just-in-time compilation and prototype pollution that challenge current static analysis approaches.
\end{itemize}

\section{Conclusion}
The proliferation of web applications has intensified security risks from malicious JavaScript, where advanced obfuscation techniques and dynamic language features (e.g., nested closures) challenge conventional detection methods. Our work introduces a novel defense framework that synergizes LLM-based semantic deobfuscation with hierarchical graph learning: (1) A multi-stage prompt engineering pipeline reconstructs original code semantics from obfuscated inputs, generating normalized AST representations; (2) To overcome structural noise from JavaScript's dynamic typing and scope fragmentation, we develop a Cluster-wise Graph Transformer that jointly models node-level semantics and cluster-induced relationships through innovative node-to-cluster attention. Experimental validation shows our method achieves 94.64\% and 97.71\% F1-scores (10.74\%/13.85\% absolute gains over SOTA) with 4.82×-5.91× higher TPR at critical FPR thresholds, while maintaining exceptional cross-dataset consistency (3.07\% F1 variance). This hybrid paradigm of semantic-aware LLMs and structure-preserving GNNs establishes a new foundation for robust malware detection across evolving threat landscapes.

\bibliographystyle{cas-model2-names}

\bibliography{cas-refs}


\bio{portrait/lzh}
\textbf{Zhihong Liang} holds a Master of Engineering in Computer Applications from South China University of Technology. He is a Senior Professor-level Senior Engineer, a member of the Artificial Intelligence Committee of the Chinese Society for Electrical Engineering, and a member of the Computer Security Committee of the China Computer Federation. His long-term research and engineering activities focus on network security, data security, artificial intelligence, and cloud computing. He has authored or co-authored more than ten academic papers.
\endbio

\bio{portrait/wx}
\textbf{Wang Xin} holds a Master of Engineering degree in Mechanical Design and Theory from Northeast Agricultural University. She is currently employed at Xidian University, where she is engaged in the research and development of industrial software and intelligent security.
\\
\\
\\

\endbio

\bio{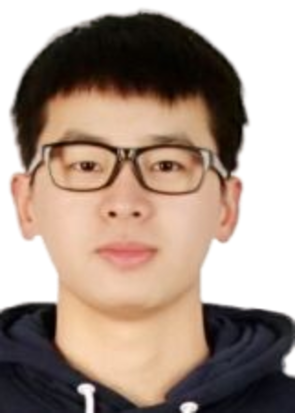}
\textbf{Zhenhuang Hu} is currently pursuing his graduate studies at the Hangzhou Institute of Technology of Xidian University. He obtained his Bachelor's degree from Harbin University of Science and Technology. His primary research interests encompass malicious code detection, network security, and applications of large language models.
\\
\\
\\

\endbio

\bio{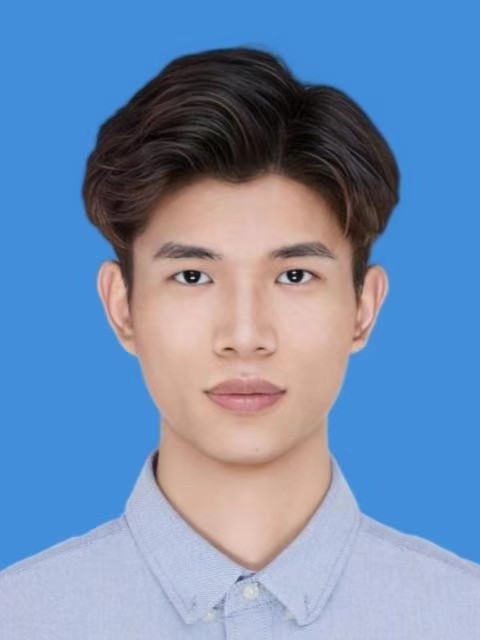}
\textbf{Liangliang Song} is a graduate student at the Hangzhou Institute of Technology of Xidian University. He received his Bachelor's degree from Hangzhou Dianzi University. His research focuses on deep learning, network security, and large language models.
\\
\\
\\

\endbio

\bio{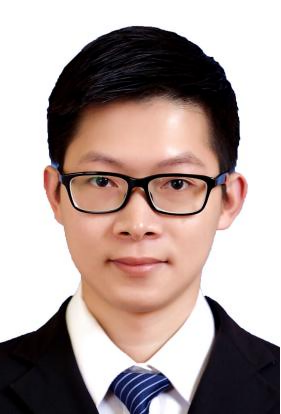}
\textbf{Lin Chen} received his M.E. in Computer Science and Technology from South China University of Technology. He is an engineer whose research centers on AI security in power systems and network security, with more than ten relevant academic papers published.
\\
\\
\\

\endbio

\bio{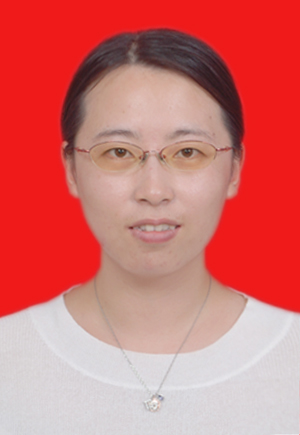}
\textbf{Jingjing Guo} received her B.S., M.S., and Ph.D. degrees from Xidian University, China. Her research focuses on AI system security, privacy protection, UAV and IoT security. She has published over 50 papers in top journals such as IEEE JSAC and TMC, and holds more than 60 patents.
\\
\\
\\

\endbio

\bio{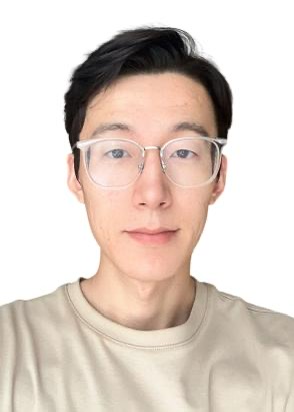}
\textbf{Yanbin Wang} is currently an Associate Professor at the Hangzhou Institute of Technology, Xidian University. He received his Ph.D. in Cybersecurity from Zhejiang University. His research focuses on AI-powered web security, software security, and blockchain security. Dr. Wang has led or participated in over 10 key research projects, including National Key R\&D Programs and provincial-level initiatives. With 60+ publications in top-tier international journals and conferences, he previously served as an editorial board member for three SCI-indexed journals.
\endbio

\bio{portrait/ty}
\textbf{Ye Tian} is an associate professor at Xidian University. He received
Ph.D. from Harbin Engineering University. His current research interests include artificial intelligence security and multimodal information processing.
\endbio

\end{document}